# SCC-YOLO: An Improved Object Detector for Assisting in Brain Tumor Diagnosis

Runci Bai
China Academy of Information and Communications Technology
Institute of Cloud Computing and Big Data
Beijing, China
byimei@126.com

Guibao Xu*
China Academy of Information and Communications Technology
Institute of Cloud Computing and Big Data
Beijing, China
xuguibao@caict.ac.cn

Yanze Shi
China University of Mining and Technology
School of Information and Control Engineering
Xuzhou, Jiangsu, China
mr_yanze_shi@163.com

## Abstract

Brain tumours may cause neurologic impairment, cognitive and mental disorders, elevated intracranial pressure and convulsions, which may be associated with a serious health hazard. The You Only Look Once (YOLO) system has been demonstrated to be highly accurate when it comes to detecting objects in medical imaging. A new SCC-YOLO structure is proposed, which combines SCConv with YOLOv9. The SCConv module improves the performance of the convolutional neural network, which improves the performance of the system. In this paper, we investigate the impact of YOLOv9 on the Br35H dataset (Brain_Tumor_Dataset) Results indicate that SCC-YOLO improved mAP$_{50}$ by 0.3% on the Br35H dataset and by 0.5% on our custom dataset compared to YOLOv9. SCC-YOLO achieves state-of-the-art performance in brain tumor detection.

## 1 Introduction

Magnetic Resonance Imaging (MRI) is the most effective imaging technique for visualizing the brain and identifying tumors[1]. However, due to the varied morphology and relatively indistinct edge characteristics of brain tumor images[2], the process of diagnosing brain tumor conditions through magnetic resonance imaging (MRI) is both complex and inefficient for clinicians, resulting in an elevated risk of misdiagnosis and missed detection. Researchers have applied machine learning techniques to the segmentation and classification of brain tumor images[3-8]. In the automatic detection and auxiliary diagnosis of brain tumors, relevant researchers have applied techniques such as unsupervised learning[9], convolutional neural networks (CNN)[10], deep stacked autoencoders (DSAE)[13], and You Only Look Once(YOLO)[11], [12-16]. Dhabliya et al. applied the YOLOv3[17] model to the computer-aided detection of brain tumors in MRI scans, representing an important study of the YOLO series models in brain tumor detection[14]. Kang et al. innovatively proposed the RCS-YOLO[15] based on YOLOv7[33] and BGF-YOLO[16] models based on YOLOv8[18], achieving good accuracy and speed on the Br35H dataset[23], demonstrating the high feasibility of the YOLO series in brain tumor image detection.

The Programmable Gradient Information (PGI) idea is introduced in YOLOv9[19], and it is used to update the weight of the network by getting reliable gradient information. This method solves the problem of data losing in the process of extracting and converting the data, and achieves optimal precision and fast performance in MS COCO dataset. In order to improve the YOLOV9 model's performance, a variety of attention mechanisms have been added to the existing network architecture. The FMSD Module (Fine-grained Multi-scale Dynamic Selection Module) is presented by Yukang Huo and his colleagues, which uses a more efficient approach for selecting and fusing dynamic characteristics on fine scale multiscale characteristic maps, and AGMF Module (Adaptive Gated Multi-branch Focus Fusion Module) is used to combine all kinds of characteristics that are caught in different branches. They integrated these two modules into YOLOv9 to develop a novel object detector with higher detection accuracy[20]. Weichao Pan et al. proposed EAConv (Efficient Attention Convolution) and EADown (Efficient Attention Downsampling), and designed a lightweight model called EFA-YOLO (Efficient Feature Attention YOLO) based on these two modules. It has greatly increased the detecting precision and the reasoning speed of the fire detection application [21]. Yifan Feng et al introduced Hyper-Yolo, which is an approach that integrates multiple layers of information into a semantic space, which can be used to integrate multiple layers of information and make use of high-level features. This approach has been shown to have excellent performance on the COCO data set, which has been shown to be one of the most advanced architectures in the world[22].

*Corresponding author.

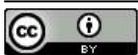

**Table 1: Data Division.**

|  | Train Set | Test Set | Total |
|---|---|---|---|
| Numbers of Images | 7920 | 1980 | 9900 |
| Numbers of label files | 7920 | 1980 | 9900 |

Therefore, in this paper, we propose a novel model named SCC-YOLO based on the YOLOv9 that introduces SCConv attention into it to further improve the detection results. Based on the research and experience, we have achieved the following results: (1) By gathering, choosing, marking, treating and cleaning brain tumour images, we have built a Brain _ Cancer Data File with a total of 9900 specimens. The RGB (Picture Resolution: 139x132 Pixels) is composed of 7 920 Training Picture Data Sets, and 1 980 Sample Picture Sets to be Developed. (2) Combining SCConv with the head of the initial original network helps better extract characteristic information regarding images of cranial tumors more excellent manners. (3) At the same time combined the SE attention mechanism together within the head section of our first original networks designs experiments comparing impacts caused among several diverse differences between selection concerning various existing kinds attention algorithmic structures applied to our unique special brain tumorous disease imagery identification problems diagnosis categories situation's detection effects. (4) To the best of our knowledge, this is the first time that the enhanced YOLOv9 has been applied to brain tumor detection.

## 2 MEthods
### 2.1 Data Preparation

We used the publicly available dataset Br35H[23] and our custom dataset Brain_Tumor_Dataset for model training and testing.

Br35H was built by Ahmed Hamada with 803 MRI pictures labeled by brain tumor, including 501 train image pictures, 202 certification picture images, and 101 test picture images. The entire architecture offers a sufficient number of reference specimens for subsequent location of tumour detection/decision-making in relation to the classification of brain tumours.

Because the Br35H data set is too small, we have developed a Brain _ Tumour _ Data Set with Labelimg. The data set consists of 9900 RGB images at 133x132 resolution, with distinct borders and full pictures, as well as associated tag txt files. This data set consists of 3 tags, called Tag 0, Tag 1, and Tag 2, which represent 3 distinct types of brain tumours. Each image is marked with multiple labels. The train set consists of 7,920 images and 7,920 label files, while the test set includes 1,980 images and 1,980 label files, as shown in Table 1.

The Brain _ Tumour _ Data Set has a much larger sample size than the current publicly available data sets, and it can be used to improve classification performance. It has a medium resolution, which can preserve the detail of the picture at the same time, and can be used in YOLO family models. Moreover, the integrity of the picture sets is helpful in eliminating the problem of losing or corrupted images and making sure that the model is able to learn from high quality data. Figure 1 illustrates some of the typical images from the dataset.

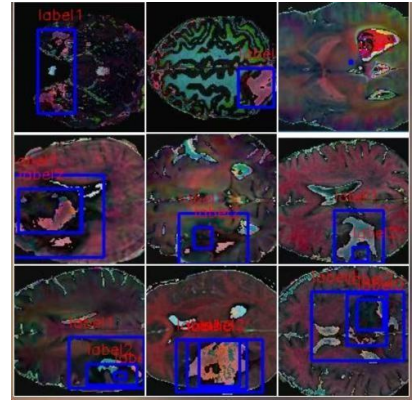

**Figure 1: Part of the dataset sample display.**

### 2.2 Overview of SCC-YOLO

In Figure 2, SCC-YOLO is presented, where SCConv [24] is inserted into the old YOLOv9 architecture, and this module is located on the thirty-seventh level of YOLOv9.

The structure is composed of two major parts: the spine and the head, which are made up of a number of carefully placed layers that help it perform as a whole.

YOLOv9 is based on deep neural networks, which is based on deep convolutional neural networks. This model is based on a deep neural network, which is composed of several convolution layers, which gradually decrease the size of the input image.

The second convolution is used to reduce the size of the output. Then, the next layer will take a step back to P2/4 and P3/8. The second convolution is used to reduce the output size by half, and then the next one is reduced to P2/4 and P3/8.

The backbone utilizes a number of RepNCSPELAN blocks. The purpose of these blocks is to improve the performance of the system by using a combination of residual links and efficient management of channels. In particular, they increase the dimension of the feature from 256-512, and keep the computation efficiency and expression ratio in equilibrium. A number of RepNCSPELAN blocks are used in the backbone. The purpose of these blocks is to achieve better performance by using a combination of residual links and carefully managing channels. Using these blocks, the characteristic dimensions are scaled from 256-512 to achieve an efficient and expressive balance.

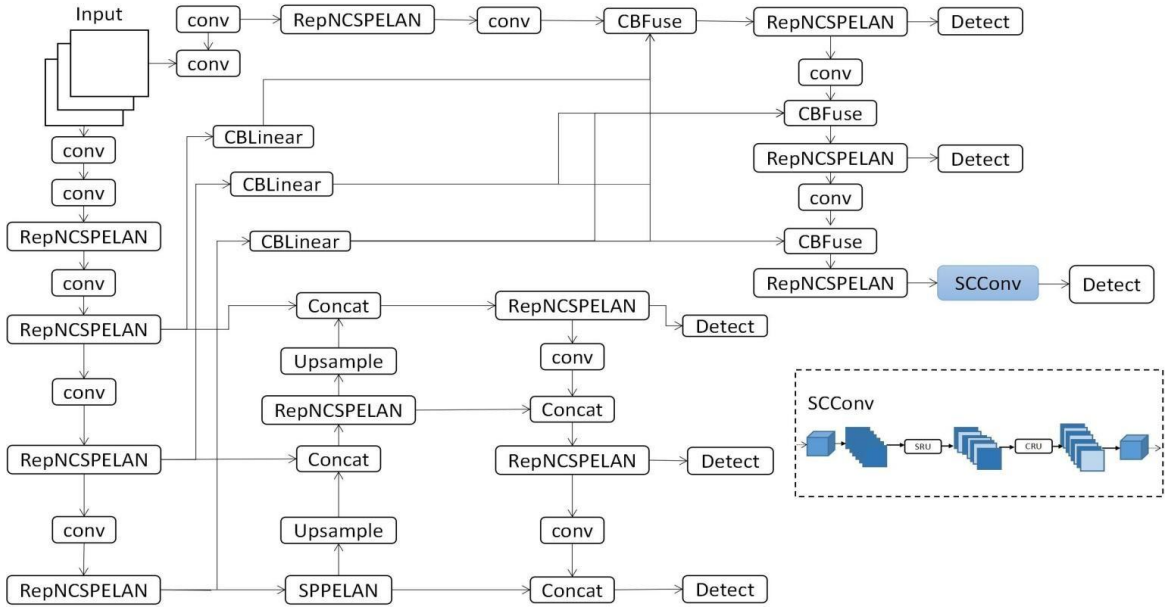

Figure 2: Shows the SCC-YOLO overall framework.

## 2.3 Intergration of SCConv

After the 37th layer of the YOLOv9 network head, we merged the SCConv module, which is a plug-and-play operation that combines SRU and CRU in series, as shown in Figure 2.

Then we refine the spatial intermediate input features via the SRU operation, and carry out the CRU operation to get the intermediate input features refined in the channel dimension. The SCConv combines the redundancy of space and channel in the intermediate feature and integrates these blocks into the backbone YOLOv9. It efficiently reduces redundancy for the intermediary feature maps. We design diagrams to explain more clearly about the architecture with Figures 3 and 4. As can be seen from Figure 3, the original information of the intermediate map is separated based on weight differences that exist between one another by means of comparing two times after being multiplied through the matrices inside SxConvd and fused back again within Sc_Convd. Figure 4 indicates the overall structure of CRU. Similarly as stated at the beginning above, we first divide channels and then apply transformation techniques along the transformed parameters (channel) lines until reaching down to the fusion section before it was depicted earlier. Once further combined all together, those divided paths become combined afterwards, resulting in achieving fewer parameters while keeping the same detection capability accuracy due to such a trick. Referring to the diagram presented with reference figure, CRU first splits out channels and then transforms them separately across tansformed columns line finally fusing together so we obtain lesser parameters along with detection capabilities remaining the same after this trick is applied earlier.

## 2.4 Comparison with SE Attention Mechanism

Among them, the common academic SE (Squeeze-and-Excitation) attention mechanism [25] has attempted to improve the model performance by focusing on enhancing the expressiveness of channel features. Once the feature channel descriptor is obtained during the'squeeze' stage, it will then proceed through the the process of adaptively adjusting the contribution ratio of different channels is such that it gives emphasis to the important features while making those that are not essential in a given input or stimulus situation less prominent. In actual practice, this entails utilizing global average pooling to get descriptive parameters for each channel, generating channel weights through full connection mapping, and finally applying these weights back to the original image to adjust the importance levels of the channels. A number of specialists and researchers have also included YOLO releases beginning with v1 in their papers [26–32]. But they only treat the channel according to the characteristic map information with an emphasis weight treatment, without considering the other information included in the space. This usually ignores the helpful space-relevant details found on certain specific types of images, for example, complicated tumour sites, which may lead to a marked difference in classification performance when combined. Furthermore, because of the above mentioned SE architecture, which includes Gobal Averaging Pooling and has an additional forward pass in each convolution layer, there is a need for further processing. Thus, although this approach has been demonstrated in a wide range of visual fields, such as classification systems, there is still a lot of scope to improve performance against more complex and more reliable solutions, particularly those that involve complex interaction between functions such as medical detection. That's why here we tested the implementation integration capability effect after layer 37 while keeping all other experiment protocols intact following our previously defined protocol framework used for SCC-YOLO training routine. It can be seen from the table comparisons reported thereafter which state the overall metrics performances that regardless of which benchmark is considered, the Br35H/Brain_Tumor_Dataset uniformly yield poorer results relative to what was achieved previously using the SCC-YOLO trained architecture version respectively, indicating an inherent drawback in the standard SE-YOLOv9 designs that they are unable to perform competitively in terms of accurate medical-grade diagnoses even though the basic functioning aspects work fine.

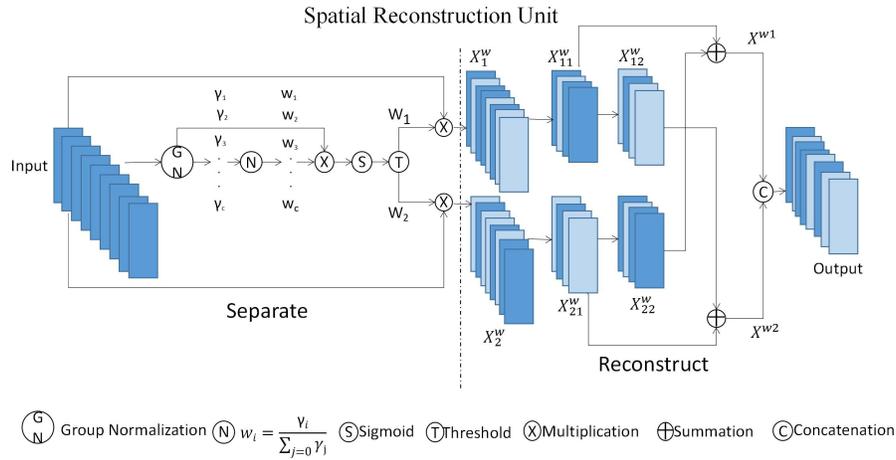

**Figure 3: The architecture of Spatial Reconstruction Unit.**

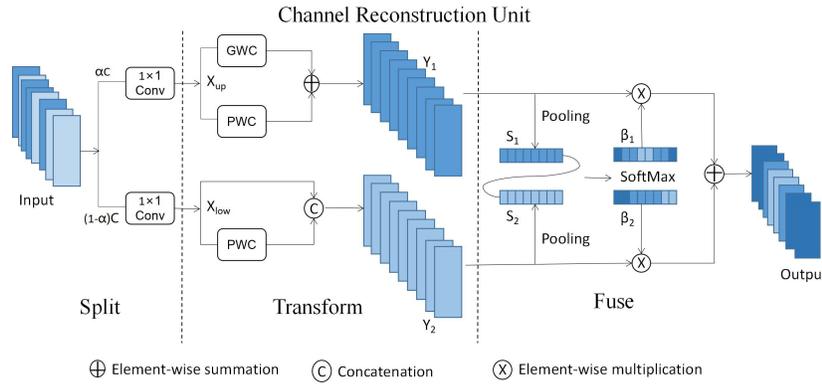

**Figure 4: The architecture of Channel Reconstruction Unit.**

## 3 Experimental Details
### 3.1 Experimental Evironment and Setup

The SCC-YOLO model was trained and tested on an NVIDIA GeForce RTX 3090 platform. In Table 2, we realized the related method research based on YOLOv9c, using different optimization algorithms to complete target detection experiments with comparable training hyperparameter configuration sets: SGD, momentum=0.937; for example, the optimizer used in comparing the SCC-YOLO model was as follows: batch_size=4, epochs=120 when tested and trained (number of times processed) on each image sample on the Br35H dataset database during network parameter learning. Similarly, according to their significant data samples differences respectively, on the large dataset with thousands upon thousands or even tens of thousands of records among other existing Brain_Tumor datasets Brain_Tumor_dataset, both methods employed in training with parameters setting are: batch-size=4, epochs=400. The optimizer is also stochastic_gradient_descent() as defined above.

### 3.2 Evaluation Metrics

In this paper, we select precision, recall, $mAP_{50}$ and $mAP_{50:95}$, parameters, layers and gradients as evaluation metrics for model performance in order to study the advantages and disadvantages of the model.

Using IoU = 0.5 as the standard, precision and recall are obtained from the following formulas:

$$Precision = \frac{TP}{TP + FP} \quad (1)$$

**Table 2: Experimental Setup**

|  | Batch_Size | Epoch | Learning Rate | Momentum | Regression Loss Function | Optimizer |
|---|---|---|---|---|---|---|
| Br35H | 4 | 120 | 0.01 | 0.937 | CIOU | SSD |
| Brain_Tumor_Dataset | 4 | 400 | 0.01 | 0.937 | CIOU | SSD |

$$Recall = \frac{TP}{TP + FN} \quad (2)$$

In this context, TP denotes the quantity of positive samples that are successfully identified as positive sample points; FP indicates the number of negative samples that are wrongly judged as positive samples; and FN represents the number of positive samples that are misjudged as negative samples.

mAP50 stands for the average value of different precision rates on positive samples (that is, in terms of PR). This is done by taking the mean of each point produced between recall and accuracy on the model's positive samples at an interval of 0.1. The second method includes, however, calculating at a number of thresholds between 0.5 and 0.95 (ten in all) steps of 0.05. This will result in a very stringent index, which is advantageous in showing the difference in the performance of the model between the test results with varying degrees of difficulty.

This entry measures the quality of the inner structure or parts of a mathematical model (such as neural networks). In general, we can calculate the total amount of parameters by adding up all the weights of each layer along with all the biases of each layer. Generally speaking, a model with more than one parameter will be more complicated, though it is possible that it will be more efficient to describe the diversity of the data set than it is for a relatively simple model.

The gradient reflects the degree of change in the loss-function-values of the current neural nets with respect to the changes that occur at certain positions located anywhere along the respective optimizing-processions that are guided in the direction pointed towards such gradients themselves when searching for the global minima. It also represents specific vectors that comprise combination product pairs consisting of partial-derivatives-of-scalar-functions versus each member variable within the corresponding coefficient-groups associated accordingly whenever training deep learning projects through algorithmic methods such as the Stochastic Gradient Descent SGD methodological strategies and so on.

## 4 Experimental Results and Discussion Analysis

Table 3. shows different model's performance metrics evaluated on the Br35H Dataset, and they are Mean Average Precision IoU equals 0.50 ($mAP_{50}$), Mean Average Precision under multiple IoUs from IoU=0.50 to IoU = 0.95 ($mAP_{50:95}$), Preciseness, and Recall.

YOLOv9 achieved $mAP_{50}$ as 0.954, $mAP_{50:95}$ as 0.751, Precenseness as 0.926, and Recall as 0.939; SE-YOLOv9 is a little behind him: he has a MAPI of 0.931, a MAP50: 95 of 0.697, a precision of 0.906 and a Recall of 0.914. However, a comparison of these three models shows that the proposed approach and YOLOv9 provide excellent results, while SE-YOLOv9 is not efficient in detecting tasks. The $mAP_{50}$ value of SCC-YOLO reached 0.957, an improvement of 0.003 compared to YOLOv9. As mentioned above, with such improved accuracy, precision parameters increased, making the SCC-YOLO slightly advantageous.

Table 4 lists the performances of three models on the evaluation dataset. The experiment results show that the YOLOv9 has a detection accuracy of $mAP_{50}$ = 0.855, which is used for benchmark learning; while obtaining an $mAP_{50:95}$=0.631, Precision=0.938, Recall=0.783. From these metrics, we can conclude that this model performs very well in detecting objects; The SE-YOLOv9's testing outcome reached $mAP_{50}$=0.828, compared with 0.855 detected by YOLOv9; meanwhile getting $mAP_{50:95}$=0.585, Precision=0.906, Recall=0.748 shows our proposed SE-YOLOv9 doesn't perform as well as YOLOv9 did when facing real-world object recognition problems; As SCC-YOLO ours surpasses SE-YOLOv9's level by gaining $mAP_{50}$ score 0.860 - up from 0.855 but just under previous result of $AP_{50}$ + 0.005, besides having more considerable effect than current SE-YOLOv9 by increasing $AP_{50:95}$+0.032 — respectively equaling $mAP_{50}$=0.633; p=0.929; r=0.781.

As can be seen from Table 5, we gave a comparison on the three aspects of parameters: numbers of total parameter;numbers of total layer; usage of gradient totally. The YOLOv9 completely has 50,999,590 parameters entirely and reaches 962 in quantities of composition totally, using quantity is 50,999,558 usage of gradient totally. The SE-YOLOv9 raises its parameter by hundreds thousands over hundred thousand and comprises altogether millions to tens than that of compared SCC-YOLO (ours). As for layers in quantity composition absolutely they reach all 934 exactly in number wholly, utilizing quantity counts up to nearly 60M precisely at gradient totally applying them. Besides, the one comparing comparatively with parameter amounts considerably larger than above two figures appears to us finally is SCC-YOLO but not surpassing too high than theirs totally which belongs comparatively lower in numbers as part composed partially within it besides possessing competitive edge over others still.

## 5 Conclusion

This paper presents a new SCC-YOLO based on the integration of SCConv and YOLOv9 network for Magnetic Resonance Imaging (MRI) detection of brain tumours. The SCConv module adds a significant reduction in space and channel redundancy, as well as facilitates efficient learning from medical images. Our experimental results have shown that proposed SCC-YOLO performs better than pure original model in MRI-based brain tumor detection task while tested on standard publicly available dataset namely Br35H and custom one named Brain_Tumor_Dataset having significant Mean Average Precision score respectively as 0.957 on the former dataset whereas 0.86 in case latter data-collection. We have achieved a slightly better mAP50 value on the Br35H dataset, with the SSC-YOLO result improving by 0.003 compared to its own benchmark. Meanwhile, on the custom dataset, the accuracy improvement reached approximately 0.005. By doing so, it makes the way toward development toward brain tumor treatment diagnosis research direction where earlier discoveries play most important role by virtue of SCC-YOLO which presently owns top of the class position between various peer algorithms competing each other for best results produced for same brain cancer identification/analyzing related queries purpose.

Table 3: Experimental Results on Br35H

| Model | mAP$_{50}$ | mAP$_{50:95}$ | Precision | Recall |
|---|---|---|---|---|
| YOLOv9 | 0.954 | 0.751 | 0.926 | 0.939 |
| SE-YOLOv9 | 0.931 | 0.697 | 0.906 | 0.914 |
| **SCC-YOLO(ours)** | **0.957** | **0.752** | 0.922 | **0.943** |

Table 4: Experimental Results on Brain_Tumor_Dataset

| Model | mAP$_{50}$ | mAP$_{50:95}$ | Precision | Recall |
|---|---|---|---|---|
| YOLOv9 | 0.855 | 0.631 | 0.938 | 0.783 |
| SE-YOLOv9 | 0.828 | 0.585 | 0.906 | 0.748 |
| **SCC-YOLO(ours)** | **0.860** | **0.633** | 0.929 | 0.781 |

Table 5: Comparison of network architectures.

| Model | Parameters | Layers | Gradients |
|---|---|---|---|
| YOLOv9 | 50999590 | 962 | 50999558 |
| SE-YOLOv9 | 60798759 | 934 | 60798727 |
| SCC-YOLO(ours) | 58080550 | 977 | 58080518 |

## Acknowledgments

The National Key Research and Development Programme of China: Ecological Technology for Inclusive Medical and Health Services (Project No. 2022YFF0903100), which provides potential funding for this study.